# Full orientation control of epitaxial MoS$_2$ on hBN assisted by substrate defects

Fu Zhang[1,2]*, Yuanxi Wang[2,3]†*, Chad Erb[1,4], Ke Wang[4], Parivash Moradifar[1,4], Vincent H. Crespi[1,2,3,5]‡, Nasim Alem[1,2]§

[1] *Department of Materials Science and Engineering, Pennsylvania State University, University Park, PA 16802, USA.*
[2] *Center for Two Dimensional and Layered Materials, Pennsylvania State University, University Park, PA 16802, USA.*
[3] *2-Dimensional Crystal Consortium, Pennsylvania State University, University Park, PA 16802, USA.*
[4] *Materials Research Institute, Pennsylvania State University, University Park, PA 16802, USA.*
[5] *Department of Physics, Department of Chemistry, The Pennsylvania State University, University Park, PA 16802, USA.*

Inversion asymmetry in two-dimensional materials grants them fascinating properties such as spin-coupled valley degrees of freedom and piezoelectricity, but at the cost of inversion domain boundaries if the epitaxy of the grown 2D layer – on a polar substrate – cannot adequately distinguish what are often near-degenerate 0° and 180° orientations. We employ first-principles calculations to identify a method to lift this near-degeneracy: the energetic distinction between eclipsed and staggered configurations during nucleation at a point defect in the substrate. For monolayer MoS$_2$ grown on hexagonal boron nitride, the predicted defect complex can be more stable than common MoS$_2$ point defects because it is both a donor-acceptor pair and a Frenkel pair shared between adjacent layers of a 2D heterostack. Orientation control is verified in experiments that achieve ~90% consistency in the orientation of as-grown triangular MoS$_2$ flakes on hBN, as confirmed by aberration-corrected scanning/transmission electron microscopy. This *defect-enhanced orientational epitaxy* could provide a general mechanism to break the near-degeneracy of 0/180° orientations of polar 2D materials on polar substrates, overcoming a long-standing impediment to scalable synthesis of single-crystal 2D semiconductors.

†yow5110@psu.edu
‡vhc2@psu.edu
§nua10@psu.edu
* These authors contributed equally to this work.

## I. INTRODUCTION

The breaking of in-plane inversion symmetry in polar two-dimensional (2D) crystals such as monolayer MoS$_2$ introduces novel physics such as coupled spin-valley degrees of freedom [1,2] and in-plane piezoelectricity [3,4]. Yet such blessings come with a curse: While the interactions of polar 2D layers with near-commensurate polar substrates are typically strong enough to disfavor arbitrary orientations and energetically favor two discrete orientations 180° apart, they are too weak to break the remaining near-degeneracy between these two orientations [5,6]. The inversion domain boundaries that then form at the lateral interfaces of merging crystallites [7,8] can degrade device performance [9] and may induce undesirable multilayer growth [10]. Such inversion domain boundaries also complicate the growth of topological insulators such as Bi$_2$Se$_3$ [11], high-$T_c$ superconductors [12], and 3D binary semiconductors [12] (even on carefully chosen lattice-matched substrates). Growth of high-quality single crystals is often associated with the discovery of new physics [13–16]; such growth outcomes have been impeded in polar 2D materials by the ubiquitous presence of inversion grain boundaries.

Prior efforts to suppress inversion domain formation include guiding lateral growth at step edges [11,12] (at the risk of inducing undesirable multilayer growth), or limiting nucleation density [10] (at the cost of slower growth rate). Interesting prior work grew transition metal dichalcogenides (TMD) directly on hexagonal boron nitride (hBN) by powder vapor transport (PVT), chemical vapor deposition [5,6,17–19], or thermal decomposition [20] to achieve scalability better than that of mechanically transferred heterostructures [21–27], but never achieved full orientational epitaxy (i.e. distinguishing inverted domains). The minimum requirement of distinguishing inversion domains in the grown TMD layer is the breaking of in-plane inversion symmetry in the substrate, limiting potential choices to layered compounds such as hBN and semiconductor surfaces such as the (0001) facets of GaN and sapphire. Here we focus on an hBN substrate due to its lack of surface inhomogeneity and dangling bonds [6]. We employ first-principles calculations to identify common intrinsic defects in the hBN substrate that can amplify the distinction between the 0° and 180° stacking geometries and enable full epitaxial growth: a paradoxical *defect-enhanced orientational epitaxy* in which structural defects (in the substrate) *improve* material quality in the layer grown above. Similar orientation control is then observed experimentally by growing MoS$_2$ on exfoliated hBN substrates using PVT, with excellent (~90%) orientational epitaxy. The geometry of the resulting population of triangular flakes is compatible with a near-seamless monolayer containing very few inversion domain boundaries. Aberration-corrected scanning/transmission electron microscopy (AC-S/TEM) confirms the atomic structure and orientation of the MoS$_2$/hBN system.

## II. STACKING DEGENERACY OF INVERSION DOMAINS

We begin by revisiting the difficulty in lifting the 0/180º near-degeneracy for TMDs stacked on commensurate or near-commensurate substrates. The local minimum energy states for MoS$_2$ stacked onto itself occurs at 0/180° interlayer orientations corresponding to the 2H and 3R polytypes with only 5 meV difference per MoS$_2$ unit [28]. The stacking orientation preference of hBN with itself is likewise weak [29]. The orientational preference of a MoS$_2$ overlayer on a hBN substrate is expected to be even weaker, given their ~28% lattice mismatch. Indeed, density functional theory (DFT) calculations performed with three different implementations of vdW corrections (DFT-D3 [30], DFT-TS [31], and vdW-DF2 [32]) in a periodic-approximate supercell that contains a 4×4 (5×5) supercell of MoS$_2$ (hBN) yield a 0/180° orientational preference of at most 0.5 meV per MoS$_2$ unit (see Appendix A for details), where the stacking with reversed bond polarities (defined by elemental electronegativities, see Fig. 1a) is only slightly preferred. This near-degeneracy is not surprising, since each atom in one layer systematically samples a variety of local environments in the other layer across their interface (Fig. 1a). While this energy difference can be made significant given sufficient area, the energy *barrier* across intermediate orientations between 0° and 60° (60° is symmetry-equivalent to 180°) also scales with area, and at a faster rate of 2 meV/MoS$_2$ (see Appendix A), effectively trapping the growing layer at 0° or 60°. The orientation is thus likely set when the MoS$_2$ flake is too small for the stacking energetics of its interior to overcome thermal fluctuations.

Can the spatial averaging across the supercell be broken by making some specific location(s) in the flake special? Along these lines, we first consider finite-size effects – i.e. edge effects and incomplete spatial averaging – by examining the orientational energetics of finite sulfur-passivated MoS$_2$ clusters, including those with areas smaller than the smallest possible coincident supercell and the smallest known Mo$_x$S$_y$ cluster Mo$_3$S$_{13}$ (Fig. S1). Even in these cases, a marginal preference of at most 2 meV per Mo was found. An intriguing orientation preference found in a recent work differs in that it used Mo$_6$S$_6$ clusters with unpassivated metal-terminated edges [33].

## III. DISTINGUISHING INVERSION DOMAINS BY A DEFECT COMPLEX

We next consider whether the spatial averaging (and the associated near-degeneracy) can be interrupted by a *localized structural defect* in the hBN substrate. Such defects may also act as natural nucleation sites. To find defects that can strengthen interlayer orientational coupling (i.e. correlating the polarities of hBN and MoS$_2$ more strongly), we systematically examine three types of pairwise interactions: between a MoS$_2$ point defect and pristine hBN, between an hBN point defect and pristine MoS$_2$, and between point defects in both MoS$_2$ and hBN, as tabulated in Fig. 1c. Darker colors indicate stronger pairwise binding $E_{\text{binding}} = E_{\text{pair}}^{\text{def}} - E_{\text{MoS2}}^{\text{def}} - E_{\text{hBN}}^{\text{def}} - E_{\text{adhesion}}$, where $E_{\text{adhesion}}$ is the pristine van der Waals interlayer adhesion, so that $E_{\text{binding}} = 0$ for pristine MoS$_2$ stacked on pristine hBN (top left of table). Mo$_{\text{ad}}$, S$_{\text{ad}}$, V$_S$, Mo$_S$ are respectively an Mo adatom, S adatom, S vacancy, and Mo substituting S, chosen from common MoS$_2$ defects with formation energies below 3 eV within the experimentally accessible range of sulfur chemical potentials [34]. The ↑ and ↓ symbols indicate MoS$_2$ defects on the sulfur plane away from or adjacent to the hBN layer. V$_B$, V$_N$, B$_N$, N$_B$, B$_{\text{ad}}$, N$_{\text{ad}}$, are B or N vacancies, antisite B or N (i.e. substituting N or B), and B or N adatoms respectively. We do not consider defects with higher degrees of complexity since they have higher formation energies (see Appendix B) and degrade epitaxy, as discuss later. We find the most strongly bound defects to be proximate adatom-vacancy pairs, with the 9.1 eV V$_B$+Mo$_{\text{ad}}$ binding being by far the strongest. Such combinations are *interlayer Frenkel pairs*: adatom-vacancy complexes that were originally studied for their compliance with charge neutrality and constant stoichiometry (i.e. without electron and elemental reservoirs [35]). Frenkel pairs typically appear as low-energy defect complexes in materials with large differences in cation and anion radii (to accommodate the interstitial), where they leave no detectable remnant if they recombine. By contrast, the "interstitial" in an interlayer Frenkel pair is actually an adatom that is accommodated by the van der Waals gap, and 'recombination' of the adatom on one sheet with a vacancy in a chemically distinct sheet leaves a distinguishable defect complex, as schematically shown in Fig. 1b. Since the V$_B$+Mo$_{ad}$ pair binds the strongest (Fig. 1c), we focus on it here and then show that its orientational control function generalizes to other defect pairs such as V$_N$+Mo$_{ad}$. This choice is further justified by the calculated formation energies of defect pairs [34,35] $E_{\text{pair}}^{\text{def}} - E_{\text{pristine-MoS2/hBN}} - n_i\mu_i$, where $n_i$ and $\mu_i$ are the number of $i$ atoms added or removed from the pristine heterostack and their chemical potentials, with the usual constraint from achieving thermodynamic equilibrium with pristine sheets $\mu_{\text{Mo}}+2\mu_S=E_{\text{MoS2}}$ and $\mu_B+\mu_N=E_{\text{hBN}}$. Defect pair formation energies are shown in Fig. 1d as functions of $\mu_S$ (referenced from the per-atom energy of solid α-S) and for $\mu_N$ set to the per-atom energy of N$_2$ (the nitrogen-rich limit [36]): Among the various defects in MoS$_2$, Mo$_{ad}$ (solid red) is the only defect that is stabilized when paired with V$_B$ (dashed red). (If X is an isolated MoS$_2$ defect, the X+V$_B$ binding energy needs to be stronger than V$_B$ formation energy to stabilize X+V$_B$ against X [35]). We therefore exclude other defect combinations involving e.g. S$_{ad}$ or Mo$_S$ for the present study. Defect formation energies from hybrid functional calculations are also shown in Fig. S2. Even though the Mo$_{ad}$+V$_B$ formation energy of at least 2 eV would still yield a negligible defect concentration, hBN defects should be preexisting so that the V$_B$ contribution to the formation energy need not be accounted for. The native V$_B$ in hBN before MoS$_2$ growth are expected to be out-of-equilibrium and passivated by hydrogen, since hBN samples are synthesized from hydrides and since H-passivated V$_B$ is ~7.7 eV more stable than V$_B$ [36], with a large migration barrier rendering them immobile below their annealing temperature of at least ~1000–2000 K [36]. Thus, taking the fully passivated V$_B$+3H complex as immobile (out of equilibrium), taking Mo$_{ad}$ as mobile (in equilibrium) with formation energy $E_{\text{Mo}}$, and taking their binding energy as $E_{\text{binding}} = E_{\text{VB+3H}} + E_{\text{Mo}} - 3\mu_H - E_{\text{VB+Mo}}$ (positive for Mo replacing 3H), then following the mass action law [37], one is tempted to conclude that the percentage of V$_B$ that combine with Mo$_{ad}$ is $exp[(E_{\text{binding}}-E_{\text{Mo}})/k_BT]$. Thus Mo$_{ad}$+V$_B$ pairing will approach completion as $E_{\text{binding}}$ overpowers $E_{\text{Mo}}$. However, this requirement on $E_{\text{binding}}$ can be alleviated. Just like defects can be immobilized by high migration barriers and become out of equilibrium [37], so can *defect pairs be locked by high binding energies and become out of equilibrium*. Removing each H and Mo from V$_B$ requires 2.3–2.7 eV and 9.1 eV respectively, so if unbinding occurs at 1000 K, our MoS$_2$ growth temperature, it would occur at rates of 2–200 s$^{-1}$ and 10$^{-32}$ s$^{-1}$. Therefore as along as $E_{\text{binding}}$>0, Mo$_{ad}$ will irreversibly replace H due to the much longer timescale of its unbinding. Indeed, $E_{\text{binding}}$ = 9.1–7.7=1.4 eV for Mo$_{ad}$.

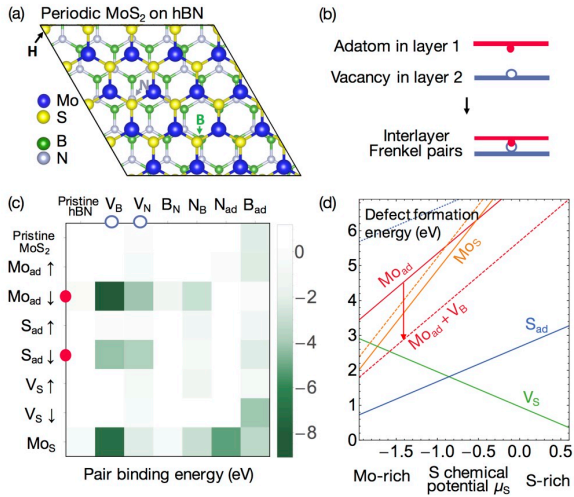

FIG. 1. (a) Top view of pristine MoS$_2$ on hBN, where S atoms sample a variety of local environments, eclipsing B(oron), N(itrogen), or H(ollow) sites. (b) Stable defect pairings in a 2D heterostack are likely Frenkel pairs: an adatom in one layer (red filled) binding strongly to a vacancy in the other layer (blue empty). (c) The Mo$_{\text{ad}}$+V$_B$ complex has the strongest defect pair binding energy (notation described in main text). (d) Formation energies of MoS$_2$ defects isolated in a monolayer (solid lines), paired with V$_B$ (dashed), and paired with V$_N$ (dotted), as function of sulfur chemical potential and in a nitrogen-rich setting.

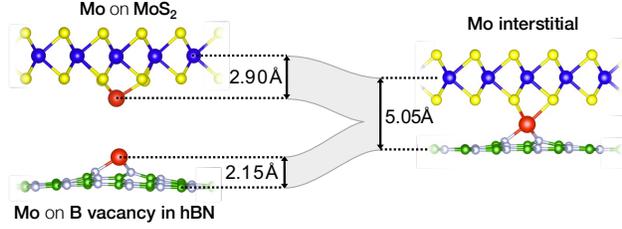

FIG. 2. A Mo interstitial atom (red) between MoS$_2$ and a V$_B$ in hBN in a 4×4/(5×5) supercell equilibrates to a 5.05 Å interlayer spacing, which is very close to the 4.96 Å spacing of pristine MoS$_2$ on pristine hBN. The individual separations of Mo from each of these sheets in isolation also sum to essentially the same value. Thus Mo+V$_B$ on hBN can nucleate the growth of a MoS$_2$ overlayer with surprisingly little deformation of the ideal bilayer spacing.

The earliest event in the formation of V$_B$+Mo$_{ad}$ is presumably the binding of a Mo atom to a V$_B$ (V$_B$ are common in hBN [38]) by 9.6 eV, consistent with the reported strong binding between V$_B$ and transition metal atoms in general [39] and the strong binding of transition metal atoms to pyridinic-nitrogen defects in graphene in particular [40] (structurally similar to V$_B$). The under-coordinated Mo atoms available in partially decomposed

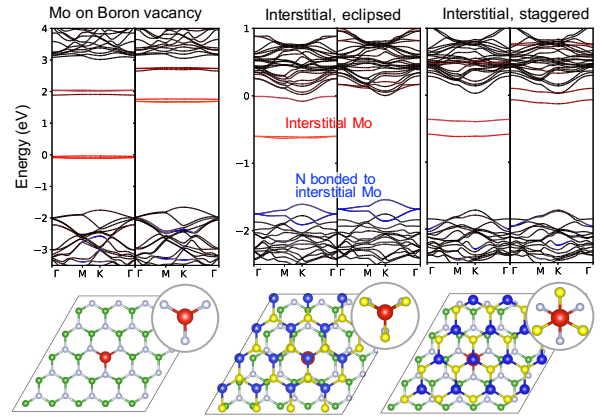

FIG. 3. Spin-polarized DFT band structures of a Mo atom bound to a V$_B$, and a Mo$_{ad}$+V$_B$ complex with an eclipsed and staggered configuration; Fermi levels are set to zero. The two band structures in each panel represent the majority and minority spin channel. States localized on the interstitial Mo and its three nearest-neighbor N atoms are colored red and blue respectively. Nitrogen levels in the valence band rise in energy when eclipsed, reflecting the repulsion between the N and S atoms bonded to the interstitial Mo.

MOCVD precursors such as $Mo(CO)_x$ or CVD precursors such as $MoO_x$ or $MoS_xO_y$ should also bind strongly to $V_B$ (see Ref. [41] for the case of $MoO_3$). The full growth kinetics for the nucleation of $MoS_2$ at a $V_B$ is beyond the scope of the present study, but the most plausible such route begins with the $V_B$-bound Mo adatom first coordinating to ambient S. Strikingly, these sulfur atoms can then form the $V_B+Mo_{ad}$ interlayer Frenkel pair by incorporating directly into a $MoS_2$ overlayer that sits above the hBN layer. Fig. 2 shows this configuration with a structurally relaxed 4×4 $MoS_2$ on 5×5 hBN supercell: the 5.05 Å interlayer separation is very close to the 4.96 Å van der Waals separation of *pristine* $MoS_2$ on hBN. Thus the Mo interstitial above $V_B$ essentially "takes up no space" in the interlayer gallery. In a further interesting coincidence, the adatom heights of two "constituent" systems – Mo above $V_B$+hBN (2.15 Å) and Mo above pristine $MoS_2$ (2.90 Å) – sum to nearly the same value.

The energetic comparisons between the 0/180º stacking described earlier are now reexamined – now including a $V_B+Mo_{ad}$ complex – with very different results. The orientation where the three sulfur atoms and three nitrogen atoms nearest to the Mo interstitial are staggered is strongly favored, by 0.88 eV per Mo interstitial, over the opposite orientation where they are eclipsed (Fig. 3). A similar preference is well-known in the conformational isomers of molecules such as ethane [42]. This defect-mediated orientational preference appears to be generic, as we also found substantial (~0.5 eV) orientational preferences for $V_N+Mo_{ad}$ and other defect-pair structures (see Fig. S3). Finally, to demonstrate the absence of local minima at other intermediate orientations and the robustness of this orientation preference against edge effects, we examined *finite* $MoS_2$ triangles on hBN with interstitial $V_B+Mo_{ad}$ at the centers and again found a substantial preference of ~0.5 eV, as shown by the connected black dots in Fig. 4 (details are discussed in the SM). The much weaker variation in stacking energy of the same flake on hBN without $V_B+Mo_{ad}$ is shown in the scattered plots, where the center of the flake lies above a B (red squares), N (blue diamonds), or hollow site (green triangles) as the flake is rotated.

Electronic structure calculations reveal the origin of the strong binding of $V_B+Mo_{ad}$ and its mechanism of orientation control: the interlayer Frenkel pair is also a donor-acceptor pair. A $V_B$ accepts three electrons from a transition metal (e.g. Mo) upon adsorption [39], leaving three degenerate occupied Mo $d$ orbitals within the band gap, as shown by the occupied red bands in Fig. 3 (the two columns in each panel are for the majority and minority spin channel). When a $MoS_2$ layer is added, these mid-gap states split differently for the two stacking orientations, but with similar summed band energies. In contrast, the eigenvalues for the orbitals of the nitrogen atoms bonded to the Mo interstitial lie much higher for the eclipsed geometry, due to the repulsion (with possible electrostatic and steric contributions [42,43]) from the sulfur above (blue bands in Fig. 3). This effect has been verified with hybrid functional calculations (Fig. S4), which generally provide more accurate defect level positions and formation energies [34,35]. The orientation preference does not extend to bilayer $MoS_2$ with a Mo interstitial, which does not charge transfer to either sheet.

### IV. GROWTH EXPERIMENTS ON PRISTINE AND PLASMA-TREATED HEXAGONAL BN

Taken in total, these results demonstrate how $V_B+Mo_{ad}$ and similar defects could induce epitaxial growth of $MoS_2$ with full orientation control. Is this mechanism borne out by experiment? To this end, $MoS_2$ was grown on freestanding hBN (on a TEM grid) as well as on $Si/SiO_2$-supported hBN using a PVT growth protocol that prioritizes the initial heterogeneous nucleation of metal species at the boron vacancy sites (see Ref. [41] and Fig. S5 for details). Raman and photoluminescence spectroscopy of this $MoS_2$ grown on hBN are similar to those of free-standing $MoS_2$, verifying the quality of the hBN substrate (Fig. S6, in contrast to $MoS_2$ on $Si/SiO_2$). Within the triangular $MoS_2$ flakes revealed by scanning electron microscopy in Fig. 5a (with more images in Fig. S8), ~90% have a single, consistent orientation in the upper region of the hBN substrate. The 0°/180° stacking degeneracy *is nearly fully lifted*. Such flakes can merge into a monolayer film nearly free of inversion domain boundaries, as suggested by ADF-STEM images in Fig. S9. A correlation between triangle orientation and the hBN surface polarity is also the most parsimonious explanation for the observed *reversal* of triangle orientation across a step edge in the h-BN substrate (dashed line in Fig. 5a), noting that the layer polarity of AA′-stacked hBN reverses across an odd-layer number step edge. Although a direct measurement of step height is not available due to its coverage by multilayer $MoS_2$ and measurement uncertainty in estimating bulk hBN thicknesses, any other explanation for this reversal would require that an alternative property not related to lattice polarity both change across the step edge and also control the lattice polarity of the $MoS_2$ flakes. The possibility that the observed orientation inversion reflects an inversion of the thermodynamic or kinetic Wulff shape is also unlikely since step edges do not interrupt Wulff shapes (except for possibly truncating corners) and also since it would imply abrupt spatial changes in the growth conditions, which vary continuously on millimeter length scales. In contrast to the clear orientation preference on hBN, second-layer $MoS_2$ flakes stacked on the first-layer $MoS_2$ film (lower right of Fig. 5a) lack preferred alignment. The bright-field TEM image and corresponding selected-area electron diffraction (SAED, Fig. 5b) confirm a precise alignment of parallel zigzag edges between hBN and $MoS_2$ (see Fig. S10 for additional characterization). Unlike in Fig. 5a, both 0/180º orientations are seen in Fig. 5b because growth occurred on *both sides* of free-standing hBN.

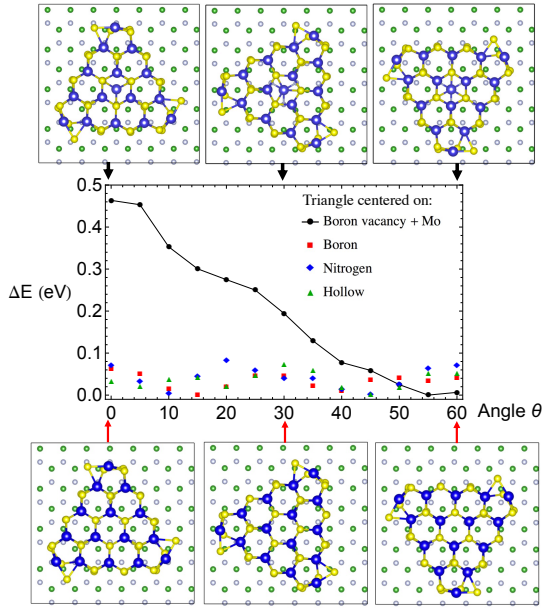

FIG. 4. Total energies (relative to the ground state for each case) of finite $MoS_2$ flakes on monolayer hBN with a boron vacancy and Mo interstitial (black, connected plot). The weaker stacking energy variations without $V_B+Mo_{ad}$ are shown in the scattered plots in color, where the center of the flake lies above a B (red squares), N (blue diamonds), or hollow site (green triangles) during rotation.

Direct imaging of single isolated boron vacancies in *multilayer* hBN substrates that are covered by $MoS_2$ is not feasible because each imaged hBN lattice site is actually a full atomic column due to the bulk hBN AA′ stacking (see Ref. [38] for a demonstration of the drastic decrease in vacancy visibility when layer number increases from one to four). While interstitial metal atoms may be more reliably imaged (as reported elsewhere for the $WSe_2$/hBN system [44]), the defect-mediated orientational control mechanism described here can be tested to a certain degree by establishing that only isolated point defects support full orientation control of $MoS_2$, i.e. more geometrically complex defects in hBN such as multivacancy voids or step edges should *not* facilitate orientational epitaxy. To test this hypothesis, a population of vacancies was introduced through a pre-growth reactive ion etching of suspended hBN films for 0, 10 or 30 seconds [41]. $MoS_2$ flakes were then grown on these plasma-treated hBN substrates with identical precursors, growth temperatures, and growth times. ADF-STEM imaging (Fig. 6a-c) along with SAED (Fig. 6d-f) reveal that plasma treatment increases the total number of $MoS_2$ flakes (likely due to a higher density of nucleation sites) while losing epitaxy, as quantified by the histograms of $MoS_2$ misorientation angles with respect to hBN in Fig. 6g-i (see also Fig. S11). High-resolution electron microscopy images (Figs. 6j-l) confirm that ion-irradiated hBN contains a much higher defect population with higher complexity, including many larger-scale voids and associated step edges, consistent with the loss of the stagger/eclipse mechanism around these more geometrically complex defects. Fresh hBN step edges created by etching should significantly promote the growth of MoS$_2$ flakes with random orientations, as suggested by the observed random orientations of MoS$_2$ flakes grown at pre-existing step edges (from the hBN sample without plasma treatment, Fig. S12).

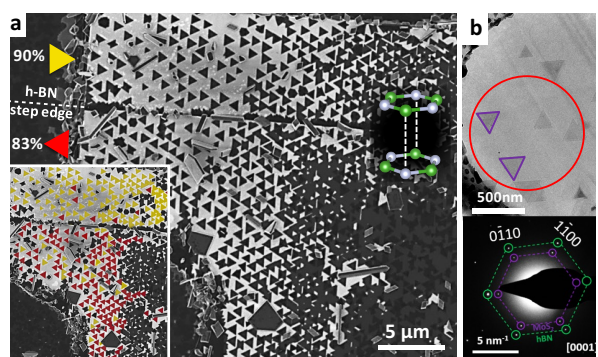

FIG. 5. (a) SEM image of triangular $MoS_2$ flakes on hBN. An hBN step edge separates two regions, each with 83% or 90% of the flakes at the same orientation. Inset shows the same image color-coded by orientation. (b) TEM image of triangular $MoS_2$ flakes grown on freestanding hBN where its crystallinity and alignment with the hBN substrate are verified by the selected area electron diffraction from the circled area.

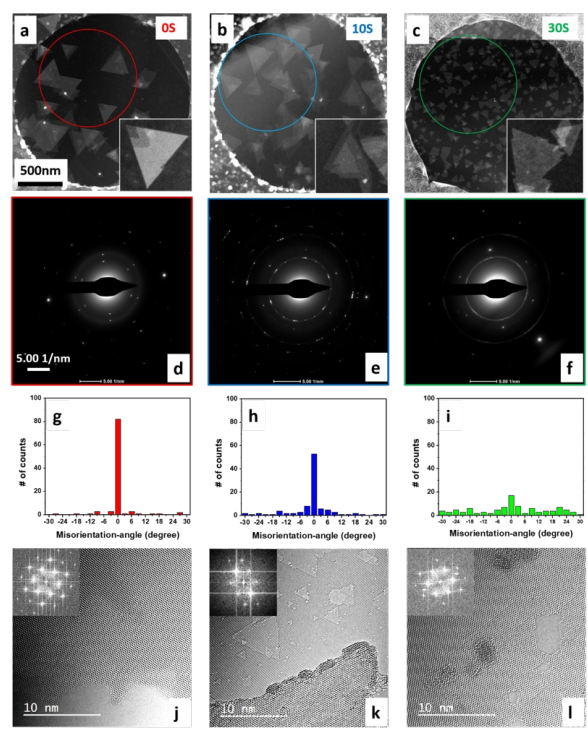

FIG. 6. Effect of different reactive ion etching time. (a-c) ADF-STEM images of as-grown $MoS_2$/hBN heterostructures after 0, 10 or 30 seconds of etching, whose degree of epitaxy is examined by (d-f) selected area diffraction, yielding (g-i) histograms of $MoS_2$ misorientation angles with respect to hBN. Corresponding (j-l) HRTEM images of hBN substrates for different etching times show more complex defect structures in the etched films.

## V. CONCLUSION

The present work demonstrates that, although vacancies in a crystal are an obvious degradation of *translational* order, their spatially "sharp" physical nature and well-controlled angular structure can paradoxically enhance the sensitivity of a system to *orientational* order, especially during the critical stage of nucleation, by accentuating orientation-dependent interlayer interactions. Defect-assisted orientational epitaxy exploits the identical structure and orientation of a given type of point defect (e.g. V$_B$) across a polar crystalline substrate. Even given full orientation uniformity and coalescence, *translational* mismatch is still a concern upon the merging of two grains. However, no such boundaries have been reported for TMDs thus far, presumably due to being outcompeted energetically by perfect stitches (see Fig. S9 and Ref. [45]). If there are no strong substrate registry effects (e.g. TMDs on hBN), the strain energy distributed deep into the flake interior across a lateral distance $D$ from the boundary scales as $D(1/D)^2 = 1/D$, so stitching is more favorable than grain boundary formation for large $D$ (i.e. large-enough flakes). To our knowledge, the only report of zero-tilt boundaries in a 2D material so far is for graphene on high-registry Ni [46]. Even misoriented grains almost always stitch together tightly into dense mirror boundaries (a chain of rhombi [9,45]), underlining the propensity for film coalescence in these systems. One can thus envision defect-enhanced epitaxy (also possibly seed molecules [17]) as providing a general means to promote well-oriented layer-by-layer growth of 2D heterostructures. These insights into the atomistic mechanisms of orientation control can help guide further improvements to film crystallinity, as has been recently achieved in the growth of WSe$_2$ on hBN using MOCVD with a strong suppression of inversion domains [44]. For example, introducing transition metal precursors of the same kind as the parent film can minimize the trapping of competing precursors that may otherwise 'poison' substrate vacancies. Coalescence techniques [47] can then be combined with orientational control to achieve monocrystallinity.

## ACKNOWLEDGEMENTS

F. Zhang and N. Alem acknowledge support from NSF under EFRI 2-DARE Grant 1433378. Y. Wang and V. H. Crespi acknowledge the National Science Foundation Materials Innovation Platform Two-Dimensional Crystal Consortium under DMR-1539916 and XSEDE (TG-DMR170050) for the allocation of computational resources on the LSU superMIC cluster. We also gratefully acknowledge the Center for 2-Dimensional and Layered Materials (2DLM) at the Pennsylvania State University. P. Moradifar and N. Alem acknowledge support from MRSEC under NSF DMR-1420620. The authors are grateful to Professor Joan Redwing for use of the APCVD system.

## APPENDIX A: ORIENTATION PREFERENCE OF PRISTINE PERIODIC STRUCTURES

In 5×5 hBN + 4×4 MoS$_2$ supercells, the relative energies of different stacking orientations and translations are calculated with three implementations of vdW corrections and are shown in Fig. 7. The three implementations agree that energies are not sensitive to translation (as shown by the clustering of the dots at 0° and 60° respectively), while the orientation preference increases from 0.1 meV (per MoS$_2$) for vdW-DF2 to 0.3 meV for DFT-D3, and to 0.5 meV for DFT-TS. Alternatively, a √21×√21 h-BN supercell (5**a**+1**b**) and a √13×√13 MoS$_2$ supercell (4**A**+1**B**) can be used to construct a heterostructure with strain less than 1% [48], where **a** and **b** are the lattice vectors for h-BN and **A** and **B** are for MoS$_2$. Since both supercell lattice vectors (5**a**+1**b**) and (4**A**+1**B**) lie about 15° degrees away from the zigzag direction, the same heterostructure supercell can fit stacking geometries close to 0°, 30°, and 60° (more accurately, 3°, 25°, 35° and 57°). Thus the two near-ground-state stackings (3° and 57°) can be fairly compared with the two intermeditate twist angles (25° and 35°), i.e. with the remaining 1% artifical strain cancelled out when comparing relative energies. The energy difference between 3° and 57° is 0.4 meV per MoS$_2$ unit, consistent with the estimate using the 5×5+4×4 heterostructure supercell, while near-30° stackings are about 2 meV per $MoS_2$ unit above the 0° ground state. This barrier should span over a wide range between 0° and 60° [49], implying that, if edge effects are ignored, a flake would translate and rotate on the substrate with negligible corrugation until it is trapped by a 0° or 60° stacking.

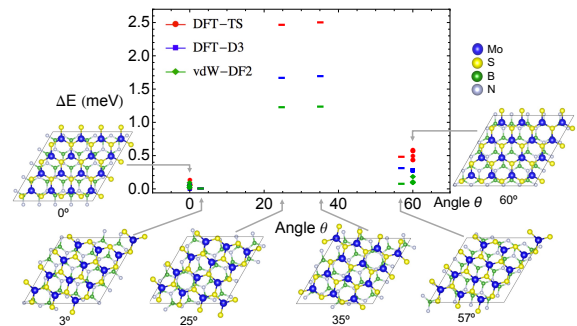

FIG. 7. Relative energies (per MoS$_2$ unit) of different stacking orientations and translations calculated with three implementations of vdW corrections. Point markers are computed with 5×5+4×4 supercells and bar markers are computed with √21×√21+√13×√13 supercells. The energy difference between the two stable stacking orientations of 0° and 60° is small compared with the barrier separating them (at intermediate twist angles).

## APPENDIX B: VACANCY TYPES IN HEXAGONAL BN

To determine whether intrinsic hBN defects with complexities higher than monovacancies need to be considered, we calculated defect formation energies of $V_B$, $V_N$, divacancy $V_{BN}$, their various hydrogen passivated complexes, and sulfur substitution of nitrogen $S_N$, as functions of the nitrogen chemical potential $\mu_N$ and the Fermi level (for charged defects) within density functional theory. Calculation methods closely follow prior studies with similar results [36,50–52], where potential alignment for the correction of spurious electrostatic interactions in supercell calculations is performed following the Freysoldt-Neugebauer-Van de Walle scheme [53] as implemented in Ref. [54]; parameters for the model dielectric profile of hBN follow those of Ref. [55] where the in-plane and out-of-plane dielectric constant of the hBN slab is properly defined. The correction energies for various supercell sizes and charged states are shown in Fig. 8a where each extrapolation towards $N_{super} \to \infty$ (using the functional form of Ref. [54]) is set to zero. The final correction energies for the $N_{super} \times N_{super} \times 1 = 5 \times 5 \times 1$ supercell geometry we used are +0.55, +2.18, and +4.90 eV for $q=\pm1$, $\pm2$, and $\pm3$. The experimentally accessible $\mu_N$ is limited within $\mu_N = E(N_2)$ and $\mu_N = E(hBN) - E(\alpha\text{-Boron})$, corresponding to N-rich and B-rich conditions. The chemical potentials for hydrogen and sulfur are set to $E(H_2)$ and $E(\alpha\text{-Sulfur})$. As shown in Fig. 8b, both hydrogen passivated $V_B$ and $V_N$ are favored against $V_{BN}$ over a wide Fermi energy range under N-rich and B-rich conditions. For unpassivated $V_B$ and $V_N$, at least one is favored against $V_{BN}$ over the same range. Thus the hBN substrate likely hosts a predominate population of the most favorable *monovacancy* point defect, each serving as a nucleation site for $MoS_2$, with consistent orientations. These results also reflect a strong binding between sulfur and $V_N$ into $S_N$ (similar to the highly stable $O_N$ impurity in Ref. [36]), since its +1 charged state is isoelectronic to pristine hBN. The strong $S$-$V_N$ binding and Mo-$V_B$ binding (see discussion in main text) are consistent with the STEM image in Ref. [44] revealing transition metal and chalcogen atoms always trapped at different sublattices of hBN.

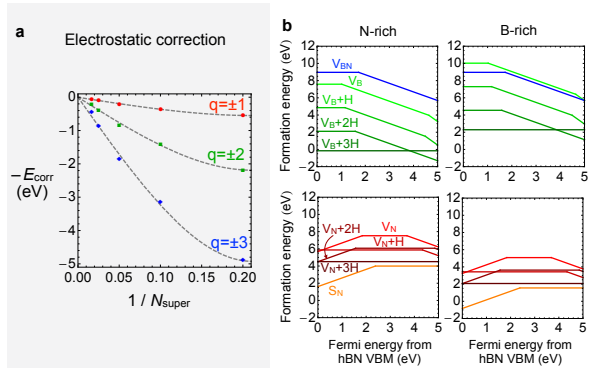

FIG. 8. (a) Corrections to spurious electrostatic interactions in supercell geometries of $N_{super} \times N_{super} \times 1$, for charged states of $q=\pm1$, $\pm2$, and $\pm3$. (b) Formation energies of $V_B$, $V_N$, and $V_{BN}$, as a function of the Fermi level at two nitrogen chemical potentials. Both $V_B$ and $V_N$ are stabilized by H-passivation and are favored against $V_{BN}$ over a wide Fermi energy range.

[1] H. Zeng, J. Dai, W. Yao, D. Xiao, and X. Cui, Nat. Nanotechnol. **7**, 490 (2012).
[2] K. F. Mak, K. He, J. Shan, and T. F. Heinz, Nat. Nanotechnol. **7**, 494 (2012).
[3] W. Wu, L. Wang, Y. Li, F. Zhang, L. Lin, S. Niu, D. Chenet, X. Zhang, Y. Hao, T. F. Heinz, J. Hone, and Z. L. Wang, Nature **514**, 470 (2014).
[4] H. Zhu, Y. Wang, J. Xiao, M. Liu, S. Xiong, Z. J. Wong, Z. Ye, Y. Ye, X. Yin, and X. Zhang, Nat. Nanotechnol. **10**, 151 (2014).
[5] M. Okada, T. Sawazaki, K. Watanabe, T. Taniguch, H. Hibino, H. Shinohara, and R. Kitaura, ACS Nano **8**, 8273 (2014).
[6] A. Yan, J. Velasco, S. Kahn, K. Watanabe, T. Taniguchi, F. Wang, M. F. Crommie, and A. Zettl, Nano Lett. **15**, 6324 (2015).
[7] W. Auwärter, M. Muntwiler, J. Osterwalder, and T. Greber, Surf. Sci. **545**, L735 (2003).
[8] F. Orlando, P. Lacovig, L. Omiciuolo, N. G. Apostol, R. Larciprete, A. Baraldi, and S. Lizzit, ACS Nano **8**, 12063 (2014).
[9] W. Zhou, X. Zou, S. Najmaei, Z. Liu, Y. Shi, J. Kong, J. Lou, P. M. Ajayan, B. I. Yakobson, and J.-C. Idrobo, Nano Lett. **13**, 2615 (2013).
[10] K. Kang, S. Xie, L. Huang, Y. Han, P. Y. Huang, K. F. Mak, C.-J. Kim, D. Muller, and J. Park, Nature **520**, 656 (2015).
[11] N. V. Tarakina, S. Schreyeck, M. Luysberg, S. Grauer, C. Schumacher, G. Karczewski, K. Brunner, C. Gould, H. Buhmann, R. E. Dunin-Borkowski, and L. W. Molenkamp, Adv. Mater. Interfaces **1**, 1400134 (2014).
[12] D. G. Schlom and J. S. Harris, in *Mol. Beam Ep.* (Elsevier, 1995), pp. 505–622.
[13] K. I. Bolotin, F. Ghahari, M. D. Shulman, H. L. Stormer, and P. Kim, Nature **462**, 196 (2009).
[14] C. R. Dean, A. F. Young, P. Cadden-Zimansky, L. Wang, H. Ren, K. Watanabe, T. Taniguchi, P. Kim,

[14] J. Hone, and K. L. Shepard, Nat. Phys. **7**, 693 (2011).
[15] X. Du, I. Skachko, F. Duerr, A. Luican, and E. Y. Andrei, Nature **462**, 192 (2009).
[16] Y. Zhang, Y.-W. Tan, H. L. Stormer, and P. Kim, Nature **438**, 201 (2005).
[17] X. Ling, Y.-H. Lee, Y. Lin, W. Fang, L. Yu, M. S. Dresselhaus, and J. Kong, Nano Lett. **14**, 464 (2014).
[18] S. Wang, X. Wang, and J. H. Warner, ACS Nano **9**, 5246 (2015).
[19] L. Fu, Y. Sun, N. Wu, R. G. Mendes, L. Chen, Z. Xu, T. Zhang, M. H. Rümmeli, B. Rellinghaus, D. Pohl, L. Zhuang, and L. Fu, ACS Nano **10**, 2063 (2016).
[20] Y. Shi, W. Zhou, A.-Y. Lu, W. Fang, Y.-H. Lee, A. L. Hsu, S. M. Kim, K. K. Kim, H. Y. Yang, L.-J. Li, J.-C. Idrobo, and J. Kong, Nano Lett. **12**, 2784 (2012).
[21] T. Georgiou, R. Jalil, B. D. Belle, L. Britnell, R. V. Gorbachev, S. V. Morozov, Y.-J. Kim, A. Gholinia, S. J. Haigh, O. Makarovsky, L. Eaves, L. A. Ponomarenko, A. K. Geim, K. S. Novoselov, and A. Mishchenko, Nat. Nanotechnol. **8**, 100 (2012).
[22] B. Hunt, J. D. Sanchez-Yamagishi, A. F. Young, M. Yankowitz, B. J. LeRoy, K. Watanabe, T. Taniguchi, P. Moon, M. Koshino, P. Jarillo-Herrero, and R. C. Ashoori, Science **340**, 1427 (2013).
[23] C.-H. Lee, G.-H. Lee, A. M. van der Zande, W. Chen, Y. Li, M. Han, X. Cui, G. Arefe, C. Nuckolls, T. F. Heinz, J. Guo, J. Hone, and P. Kim, Nat. Nanotechnol. **9**, 676 (2014).
[24] S. Larentis, J. R. Tolsma, B. Fallahazad, D. C. Dillen, K. Kim, A. H. MacDonald, and E. Tutuc, Nano Lett. **14**, 2039 (2014).
[25] O. Lopez-Sanchez, E. Alarcon Llado, V. Koman, A. Fontcuberta i Morral, A. Radenovic, and A. Kis, ACS Nano **8**, 3042 (2014).
[26] X. Cui, G.-H. Lee, Y. D. Kim, G. Arefe, P. Y. Huang, C.-H. Lee, D. A. Chenet, X. Zhang, L. Wang, F. Ye, F. Pizzocchero, B. S. Jessen, K. Watanabe, T. Taniguchi, D. A. Muller, T. Low, P. Kim, and J. Hone, Nat. Nanotechnol. **10**, 534 (2015).
[27] C. R. R. Dean, A. F. F. Young, I. Meric, C. Lee, L. Wang, S. Sorgenfrei, K. Watanabe, T. Taniguchi, P. Kim, K. L. L. Shepard, and J. Hone, Nat. Nanotechnol. **5**, 722 (2010).
[28] J. He, K. Hummer, and C. Franchini, Phys. Rev. B **89**, 075409 (2014).
[29] G. Constantinescu, A. Kuc, and T. Heine, Phys. Rev. Lett. **111**, 036104 (2013).
[30] S. Grimme, J. Antony, S. Ehrlich, and H. Krieg, J. Chem. Phys. **132**, 154104 (2010).
[31] A. Tkatchenko and M. Scheffler, Phys. Rev. Lett. **102**, 073005 (2009).
[32] K. Lee, É. D. Murray, L. Kong, B. I. Lundqvist, and D. C. Langreth, Phys. Rev. B **82**, 081101 (2010).
[33] D. Fu, X. Zhao, Y. Zhang, L. Li, H. Xu, A. Jang, S. I. Yoon, P. Song, S. M. Poh, T. Ren, Z. Ding, W. Fu, T. J. Shin, H. S. Shin, S. T. Pantelides, W. Zhou, and K. P. Loh, J. Am. Chem. Soc. **139**, 9392 (2017).
[34] H.-P. Komsa and A. V. Krasheninnikov, Phys. Rev. B **91**, 125304 (2015).
[35] C. Freysoldt, B. Grabowski, T. Hickel, J. Neugebauer, G. Kresse, A. Janotti, and C. G. Van De Walle, Rev. Mod. Phys. **86**, 253 (2014).
[36] L. Weston, D. Wickramaratne, M. Mackoit, A. Alkauskas, and C. G. Van de Walle, Phys. Rev. B **97**, 214104 (2018).
[37] C. G. Van de Walle and J. Neugebauer, J. Appl. Phys. **95**, 3851 (2004).
[38] N. Alem, R. Erni, C. Kisielowski, M. D. Rossell, W. Gannett, and A. Zettl, Phys. Rev. B **80**, 155425 (2009).
[39] B. Huang, H. Xiang, J. Yu, and S. H. Wei, Phys. Rev. Lett. **108**, 206802 (2012).
[40] Y.-C. Lin, P.-Y. Teng, C.-H. Yeh, M. Koshino, P.-W. Chiu, and K. Suenaga, Nano Lett. **15**, 7408 (2015).
[41] See Supplemental Materials.
[42] Y. Mo, Wiley Interdiscip. Rev. Comput. Mol. Sci. **1**, 164 (2011).
[43] S. Liu, J. Chem. Phys. **126**, 244103 (2007).
[44] X. Zhang, F. Zhang, Y. Wang, D. S. Schulman, T. Zhang, A. Bansal, N. Alem, S. Das, V. H. Crespi, M. Terrones, and J. M. Redwing, ACS Nano acsnano.8b09230 (2019).
[45] H. Yu, Z. Yang, L. Du, J. Zhang, J. Shi, W. Chen, P. Chen, M. Liao, J. Zhao, J. Meng, G. Wang, J. Zhu, R. Yang, D. Shi, L. Gu, and G. Zhang, Small **13**, 1603005 (2017).
[46] J. Lahiri, Y. Lin, P. Bozkurt, I. I. Oleynik, and M. Batzill, Nat. Nanotechnol. **5**, 326 (2010).
[47] X. Zhang, T. H. Choudhury, M. Chubarov, Y. Xiang, B. Jariwala, F. Zhang, N. Alem, G.-C. Wang, J. A. Robinson, and J. M. Redwing, Nano Lett. **18**, 1049 (2018).
[48] H. P. Komsa and A. V. Krasheninnikov, Phys. Rev. B **88**, 085318 (2013).
[49] D. Dumcenco, D. Ovchinnikov, K. Marinov, P. Lazić, M. Gibertini, N. Marzari, O. L. Sanchez, Y.-C. Kung, D. Krasnozhon, M.-W. Chen, S. Bertolazzi, P. Gillet, A. Fontcuberta i Morral, A. Radenovic, and A. Kis, ACS Nano **9**, 4611 (2015).
[50] S. Okada, Phys. Rev. B **80**, 161404 (2009).
[51] W. Orellana and H. Chacham, Phys. Rev. B **63**, 125205 (2001).
[52] B. Huang and H. Lee, Phys. Rev. B **86**, 245406 (2012).
[53] C. Freysoldt, J. Neugebauer, and C. G. Van de Walle, Phys. Rev. Lett. **102**, 016402 (2009).

[54] M. H. Naik and M. Jain, Comput. Phys. Commun. **226**, 114 (2018).
[55] H. P. Komsa, N. Berseneva, A. V. Krasheninnikov, and R. M. Nieminen, Phys. Rev. X **4**, 031044 (2014).
[56] J. P. Perdew, K. Burke, and M. Ernzerhof, Phys. Rev. Lett. **77**, 3865 (1996).
[57] J. P. Perdew, K. Burke, and M. Ernzerhof, Phys. Rev. Lett. **78**, 1396 (1997).
[58] D. Joubert, Phys. Rev. B **59**, 1758 (1999).
[59] P. E. Blöchl, Phys. Rev. B **50**, 17953 (1994).
[60] G. Kresse and J. Furthmüller, Phys. Rev. B **54**, 11169 (1996).
[61] T. Björkman, A. Gulans, A. V. Krasheninnikov, and R. M. Nieminen, J. Phys. Condens. Matter **24**, 424218 (2012).
[62] T. Björkman, A. Gulans, A. V. Krasheninnikov, and R. M. Nieminen, Phys. Rev. Lett. **108**, 235502 (2012).
[63] T. Bučko, S. Lebègue, J. Hafner, and J. G. Ángyán, Phys. Rev. B **87**, 064110 (2013).
[64] J. Heyd, G. E. Scuseria, and M. Ernzerhof, J. Chem. Phys. **118**, 8207 (2003).
[65] J. Heyd, G. E. Scuseria, and M. Ernzerhof, J. Chem. Phys. **124**, 219906 (2006).
[66] J. Kibsgaard, T. F. Jaramillo, and F. Besenbacher, Nat. Chem. **6**, 248 (2014).
[67] C. Attaccalite, M. Bockstedte, A. Marini, A. Rubio, and L. Wirtz, Phys. Rev. B **83**, 144115 (2011).
[68] F. Zhang, C. Erb, L. Runkle, X. Zhang, and N. Alem, Nanotechnology **29**, 025602 (2018).
[69] F. Zhang, M. A. AlSaud, M. Hainey, K. Wang, J. M. Redwing, and N. Alem, Microsc. Microanal. **22**, 1640 (2016).
[70] F. Zhang, K. Momeni, M. A. AlSaud, A. Azizi, M. F. Hainey, J. M. Redwing, L. Q. Chen, and N. Alem, 2D Mater. **4**, 025029 (2017).
[71] M. Bosi, RSC Adv. **5**, 75500 (2015).
[72] C. Lee, H. Yan, L. E. Brus, T. F. Heinz, J. Hone, and S. Ryu, ACS Nano **4**, 2695 (2010).
[73] A. Splendiani, L. Sun, Y. Zhang, T. Li, J. Kim, C. Y. Chim, G. Galli, and F. Wang, Nano Lett **10**, 1271 (2010).
[74] K. F. Mak, C. Lee, J. Hone, J. Shan, and T. F. Heinz, Phys. Rev. Lett. **105**, 136805 (2010).
[75] N. Alem, Q. M. Ramasse, C. R. Seabourne, O. V. Yazyev, K. Erickson, M. C. Sarahan, C. Kisielowski, A. J. Scott, S. G. Louie, and A. Zettl, Phys. Rev. Lett. **109**, 205502 (2012).
[76] J. C. Meyer, A. Chuvilin, G. Algara-Siller, J. Biskupek, and U. Kaiser, Nano Lett. **9**, 2683 (2009).
[77] A. Zobelli, A. Gloter, C. P. Ewels, G. Seifert, and C. Colliex, Phys. Rev. B **75**, 245402 (2007).
[78] M. Neek-Amal, J. Beheshtian, A. Sadeghi, K. H. Michel, and F. M. Peeters, J. Phys. Chem. C **117**, 13261 (2013).
[79] W. Yang, G. Chen, Z. Shi, C.-C. Liu, L. Zhang, G. Xie, M. Cheng, D. Wang, R. Yang, D. Shi, K. Watanabe, T. Taniguchi, Y. Yao, Y. Zhang, and G. Zhang, Nat. Mater. **12**, 792 (2013).
[80] A. Summerfield, A. Davies, T. S. Cheng, V. V. Korolkov, Y. Cho, C. J. Mellor, C. T. Foxon, A. N. Khlobystov, K. Watanabe, T. Taniguchi, L. Eaves, S. V. Novikov, and P. H. Beton, Sci. Rep. **6**, 22440 (2016).
[81] J. Wu, B. Wang, Y. Wei, R. Yang, and M. Dresselhaus, Mater. Res. Lett. **1**, 200 (2013).
[82] J. A. Oliveira, W. B. De Almeida, and H. A. Duarte, Chem. Phys. Lett. **372**, 650 (2003).

# Supplemental Materials for "Full orientation control of epitaxial MoS$_2$ on hBN assisted by substrate defects"

Fu Zhang[1,2,4]*, Yuanxi Wang[2,3]*, Chad Erb[1,4], Ke Wang[4], Parivash Moradifar[1,4], Vincent Crespi[1,2,3,5], Nasim Alem[1,2]

[1] *Department of Materials Science and Engineering, Pennsylvania State University, University Park, PA 16802, USA.*
[2] *Center for Two Dimensional and Layered Materials, Pennsylvania State University, University Park, PA 16802, USA.*
[3] *2-Dimensional Crystal Consortium, Pennsylvania State University, University Park, PA 16802, USA.*
[4] *Materials Research Institute, Pennsylvania State University, University Park, PA 16802, USA.*
[5] *Department of Physics, Department of Chemistry, The Pennsylvania State University, University Park, PA 16802, USA.*

* These authors contributed equally to this work.

**First-principles calculation**

Density functional theory calculations were performed using the Perdew-Burke-Ernzerhof parametrization of the generalized gradient approximation (GGA-PBE) exchange-correlation functional [1,2] and pseudopotentials constructed from the projector augmented wave (PAW) method [3,4], as implemented in the Vienna Ab initio Simulation Package (VASP) [5]. Van der Waals corrections were included using the DFT-D3 [6], DFT-TS [7], and vdW-DF2 [8] methods. Both DFT-D3 and DFT-TS show excellent agreement with random phase approximation treatments of the van der Waals interaction in the interlayer binding energy of bulk MoS$_2$ (<10% error) [9–11], but overbind hBN layers by 80–100% [10–12]. vdW-DF yields a similar binding energy for MoS$_2$ and better binding energy for hBN [10,11]. All corrections yield excellent results for corrugation, i.e. the energy variation upon sliding adjacent layers relative to each other. Ionic relaxations were all performed at the PBE level with vdW corrections using the DFT-D3 method (unless otherwise noted, e.g. for calculations using DFT-TS and vdW-DF2) until forces were smaller than 0.01 eV/Å. Hybrid functional eigenvalues and total energies were calculated using the range-separated form of Heyd, Scuseria, and Ernzerhof (HSE06) [13,14] and using structures relaxed at the HSE06 level until forces were smaller than 0.02 eV/Å.

**Orientation preference: finite flakes**

For finite triangles on pristine hBN, not only are 0 and 60° close in energy (< 2 meV), they are in fact both slightly disfavored due to edge effects, as seen from the complete orientation map of stacking energy in Fig. 4 in the main text (colored and scattered markers). This indicates that orientation control, if determined purely by van der Waals interactions between pristine sheets, can only be effective at a later stage when the triangle is sufficiently large that the substrate adhesion scaling as $L^2$ dominates over any edge effects, which scale as $L$. The near degeneracy extends down to the smallest known Mo/S cluster with a structure similar to the hexagonal motif in MoS$_2$ [15]: for this Mo$_3$S$_{13}$ cluster (Fig. S1), the orientational preference is calculated to be only 2 meV per Mo.

As with the previous series of calculations with a MoS$_2$ flake floating on h-BN, we again calculate the relative energies as the stacking orientation varies between 0 and 60°, but now in the presence of boron vacancy + Mo interstitial; these results are discussed in the main text and shown by the black connected dots in Fig. 4. Since placing Mo$_{ad}$ under a Mo site at the center of a finite triangular flake requires triangles with a side length of $l$=3 (too small to exclude corner effects) or $l$=6 (computationally too expensive, including the underlying hBN), we chose to place Mo$_{ad}$ under the hollow site with $l$=4. The corresponding periodic structure is in Fig. S3. This result is also the most direct test of orientation control for small transient clusters within the current scope of computational feasibility, since this is the earliest emergence of MoS$_2$ crystallinity that persists, i.e. the minimal prerequisite for the manifestation of a well-defined "orientation selectivity". Even if a transient oxide or oxysulfide cluster with high symmetry (e.g. C$_{3v}$) was computationally found to have a preferred orientation, the eventual MoS$_2$ flake does not necessarily inherit this orientation, since bonding characters and polarities might change.

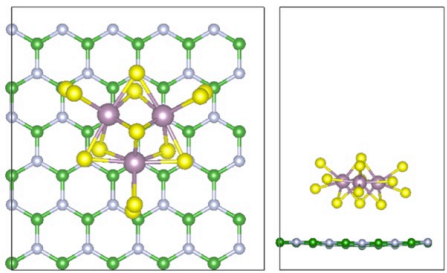

Figure S1. Mo$_3$S$_{13}$ cluster on monolayer h-BN.

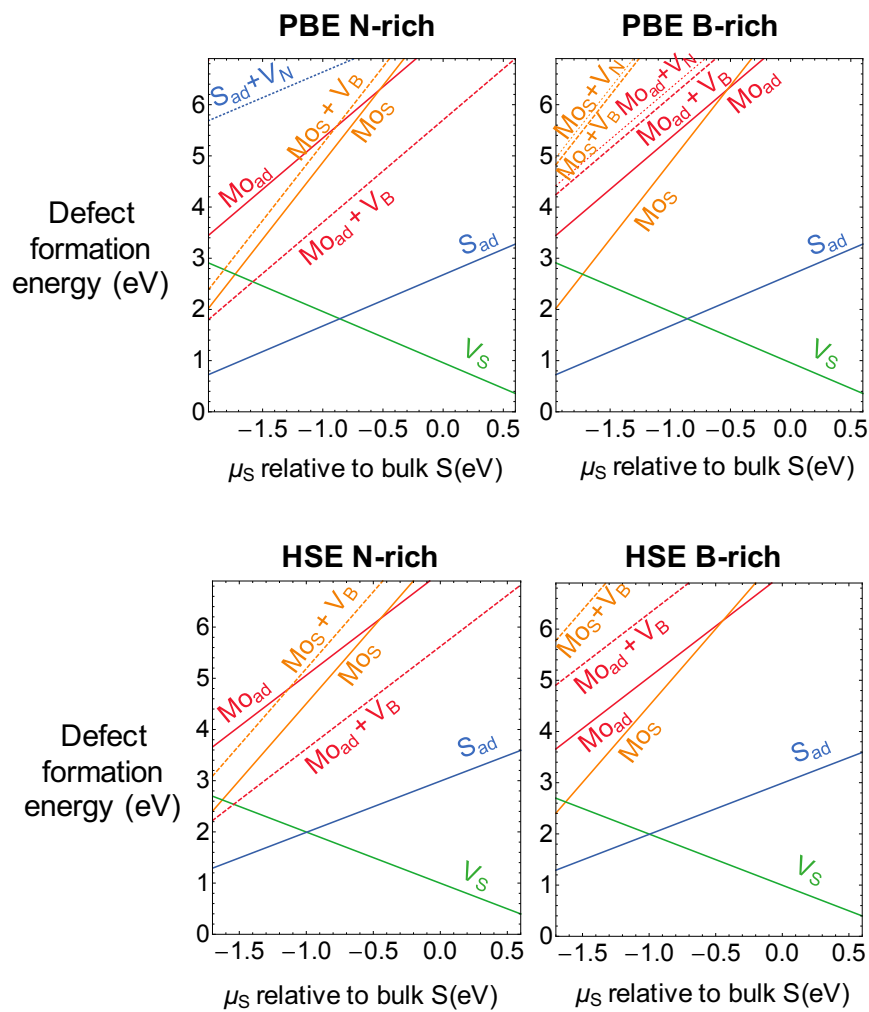

Figure S2. Defect and defect complex formation energies calculated using PBE (upper panels) and HSE06 (lower panels) in the nitrogen-rich (left) and boron-rich (right) limit. In every case the DFT-D3 method is used for van der Waals corrections. The Mo adatom is the only MoS$_2$ defect that becomes stabilized when paired with an hBN defect (a boron vacancy).

**Alternative vacancy types and Mo interstitial lateral positions**

In contrast to a boron vacancy favoring the staggered configuration over the eclipsed one, a nitrogen vacancy (Mo+V$_N$) favors the eclipsed configuration over the staggered one, by 0.49 eV, as shown in Fig. S3. This is not surprising since the boron atoms (with partial positive charge) closest to the Mo interstitial (a V$_N$ hybridizes with a Mo interstial rather than accept charge from it [16]) prefer proximity to the negatively charged sulfur atoms above. This preference, although seemingly opposite to that of a boron vacancy, would in fact not affect the orientation selectivity of MoS$_2$ flakes on hBN: a V$_N$ favoring the eclipsed configuration and a V$_B$ favoring the staggered configuration *on the same hBN layer* orient the MoS$_2$ overlayer the same.

A Mo interstial above a $V_B$ at the lateral "metal" site (i.e. under a metal atom in the $MoS_2$ layer) was investigated in the main article; for the case where the Mo interstitial is at the lateral "hollow" site, as shown in Fig. S3, the staggered configuration is still favored, by 0.53 eV. Mo interstitials at "metal" sites are presumably more likely in reality since DFT calculations show that they are always more energetically favorable than "hollow" site Mo interstitials.

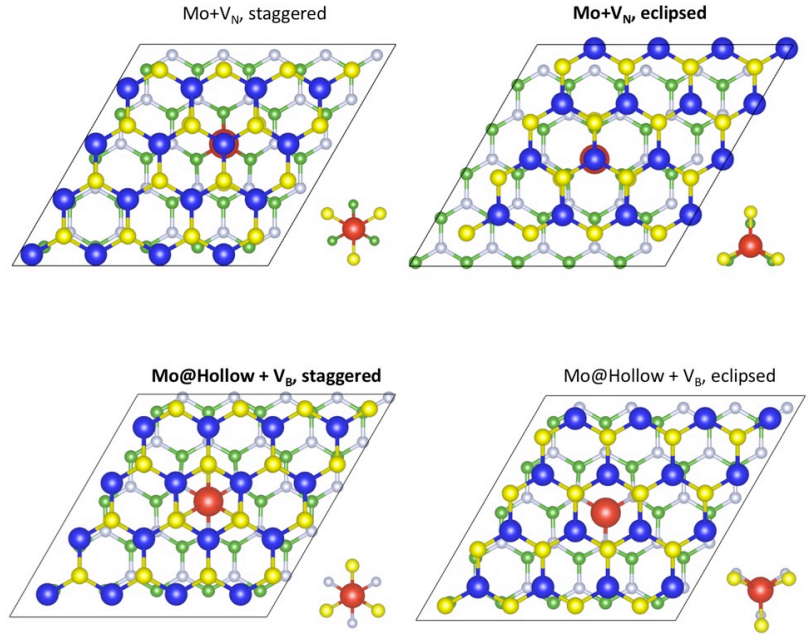

Figure S3. A nitrogen vacancy favors the eclipsed configuration by 0.49 eV. A Mo interstitial at a "hollow" site favors the staggered configuration by 0.53 eV.

**Hybrid functional calculations**

Defect levels for test cases (e.g. boron and nitrogen vacancies, not shown here) are taken from Γ-point eigenvalues using the HSE06 hybrid functional [17,18] and are verified to be consistent with previously reported values [19,20]. Defect levels for the staggered and eclipsed stacking geometries discussed in the main text (Mo interstitial at the $MoS_2$ "metal" site) are then calculated at the HSE06 level as well and compared with those calculated from PBE in Fig. S4, where the occupied defect levels are populated with spin symbols. The overall effect is an increase in band gap and defect levels shifting deeper into the mid-gap region, with no significant change in the ordering of occupied levels. The total energy of the staggered geometry calculated at the HSE06 level is lower than the eclipsed case by 0.62 eV.

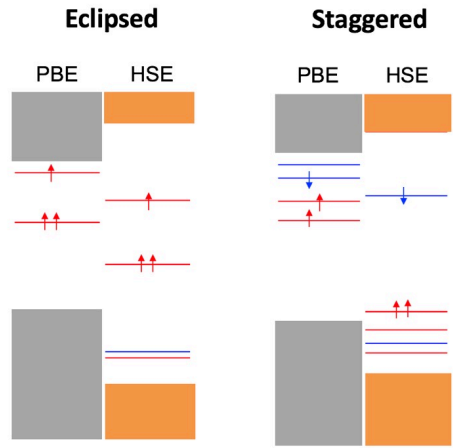

Figure S4. Defect levels for the "eclipsed" and "staggered" stacking geometries calculated from PBE and HSE06. Occupied levels are populated with spin symbols, where blue and red indicate two spin polarizations and paired spins indicate degenerate spin states.

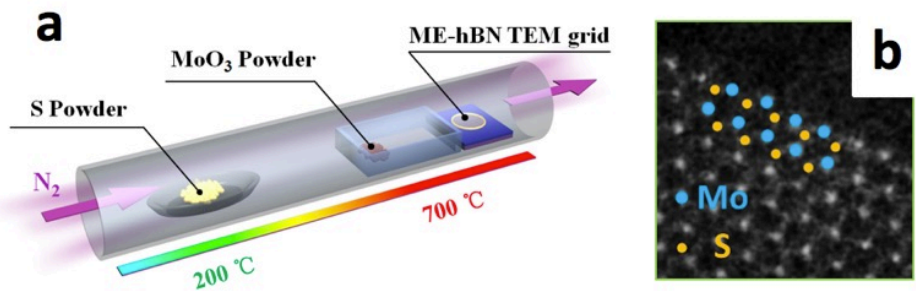

Figure S5. (a) Schematic of PVT system. (b) Annular dark field (ADF)-STEM image of a Mo-terminated $MoS_2$ edge.

**Atmosphere Pressure PVT synthesis of $MoS_2$/hBN heterostructure**

The hBN flakes were mechanically exfoliated from powder (grade PT 110, Momentive Performance Materials) and placed on a Si/$SiO_2$ substrate, before they were transferred to a Au quantifoil TEM grid using PMMA-assisted transfer [21]. The combination of mechanical exfoliation and transfer produces freestanding hBN films with high quality surfaces. Mechanically exfoliated hBN (ME-hBN) both in freestanding form and supported by a Si/$SiO_2$ substrate (300nm $SiO_2$ thickness) were used as templates to fabricate $MoS_2$/hBN heterostructures. The method of synthesizing monolayer $MoS_2$ was reported elsewhere [22]. The TEM grid was placed on a Si/$SiO_2$ substrate which follows downstream of the hot-zone crucible.

$MoS_2$ was grown on ME-hBN by powder vapor transport at atmospheric pressure in a 15mm diameter horizontal tube furnace [23], as shown in Fig. S5. High-purity nitrogen (200 sccm) was introduced into the furnace throughout the process. Approximately 1 mg of $MoO_3$ (99.99%, Sigma Aldrich) is placed in a crucible located in the hot zone of the furnace (~700°C). Sulfur powder (300mg, 99.5% Alfa Aesar) was placed upsteam of the hot zone, at a temperature of ~230°C, while the hot zone follows a ramp to 700°C over thirty minutes, a ten-minute hold at 700°C, and cool down without feedback.

**Growth protocol**

We briefly highlight two main differences between our growth protocol and that of Yu. *et al.* [24], which is operationally closest to ours among existing studies but reported a mixture of $MoS_2$ at 0 and 180º on hBN. First, while our $MoO_3$ precursor and sulfur powder temperatures are ramped up simultaneously, Yu *et al.* introduced $MoO_3$ into the cavity after S vapor filled the cavity. Exposure to S prior to Mo precursor arrival likely passivates hBN vacancies, which would disable the proposed mechanism of Mo-based orientation control. Second, Yu *et. al.* adopted a three-zone setup with typical temperatures of 115ºC (Sulfur), 450–580ºC ($MoO_3$ with delayed entry), and 750ºC (substrates), while we adopted a two-zone setup with higher $MoO_3$ temperature (i.e. higher Mo precursor partial pressure). A higher Mo precursor vapor pressure would favor direct heterogeneous nucleation of $MoS_2$ seeds on the substrate rather than the deposition of nucleated $MoS_2$ clusters from the gas phase to the substrate [25], facilitating the proposed Mo-based seeding on hBN.

**Characterization by Raman spectroscopy, photoluminescence, EDS, and EELS**

Raman spectra (Fig. S6a) reveals a frequency difference between the in-plane $E^1_{2g}$ and out-of-plane $A_{1g}$ modes that varies from 19 to 24 cm$^{-1}$ across a $MoS_2$ flake grown on BN, reflecting a variation in number of layers that is confirmed by contrast variations in the ADF-STEM image (Fig. S7) and is consistent with previous reports for PVT-synthesized $MoS_2$ crystals [26]. The monolayer regions show a photoluminescence peak at 1.89 eV (Fig. S6c), close to the optical band gap of freestanding exfoliated $MoS_2$ monolayers (1.90 eV) [27], with a similar ~50 meV full width at half-maximum [28]. The sharp, intense photoluminescence similar to that of free-standing flakes verifies hBN as an excellent substrate to preserve the intrinsic properties of the 2D sheet it supports, in contrast to $MoS_2$ grown directly on $Si/SiO_2$ (Fig. S6b).

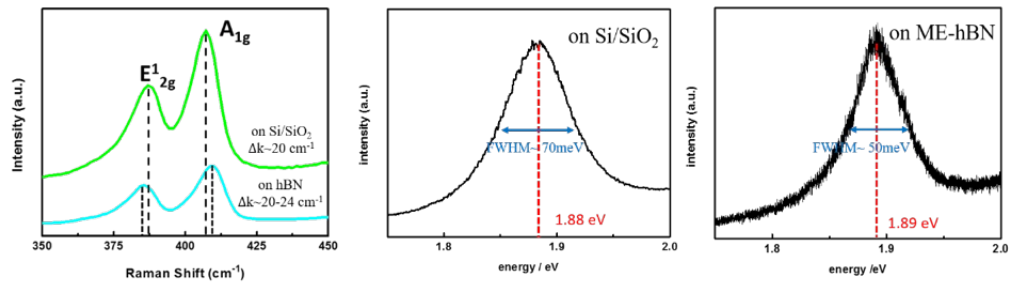

Figure S6. Raman and PL comparison between $MoS_2$ grown on hBN and directly on $Si/SiO_2$.

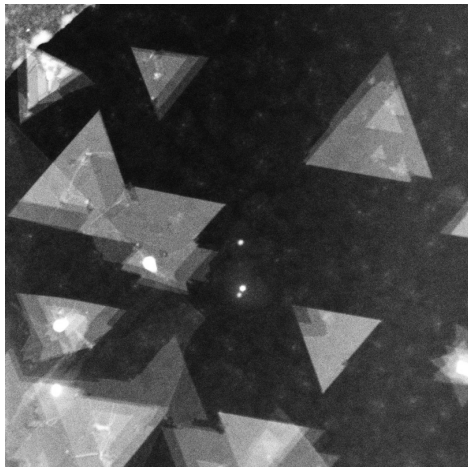

Figure S7. ADF-STEM image of triangular-shaped MoS$_2$ grown on a freestanding ME-hBN flake. Layer thickness can be identified by different contrast. The growth on the free-standing hBN can occur on both sides of the surface making both 0 and 180º degree orientations possible.

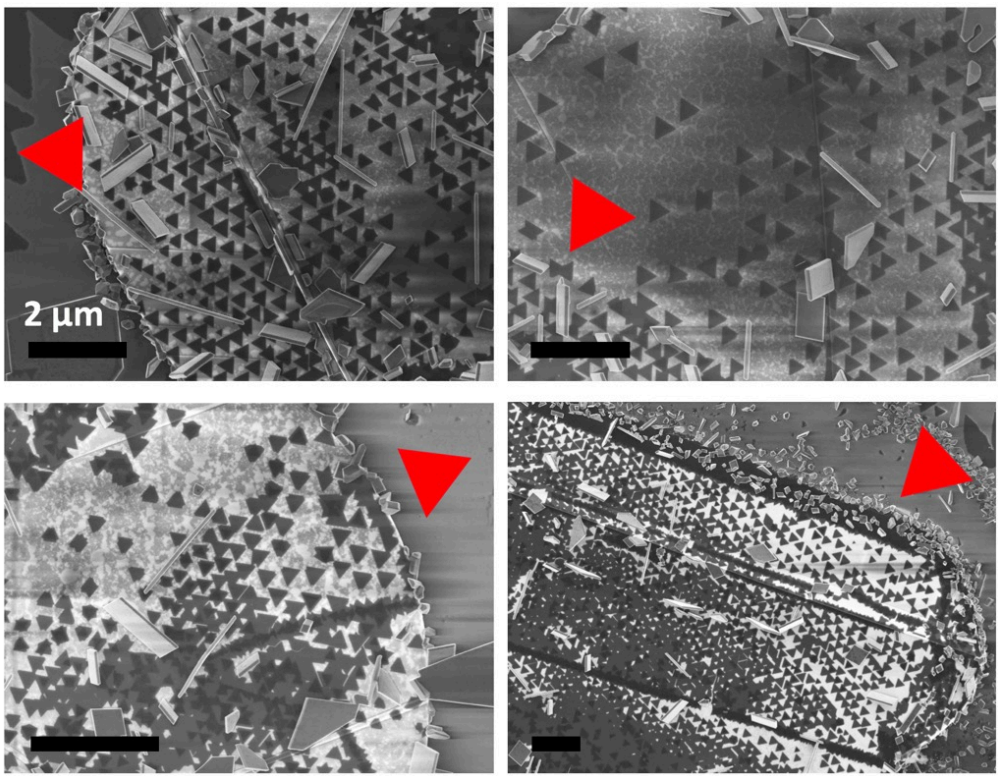

Figure S8. SEM image of more areas of triangular MoS$_2$ flakes epitaxially grown on ME-hBN on a Si/SiO$_2$ substrate (scale bar = 2 μm). Red triangles indicate the dominant orientation.

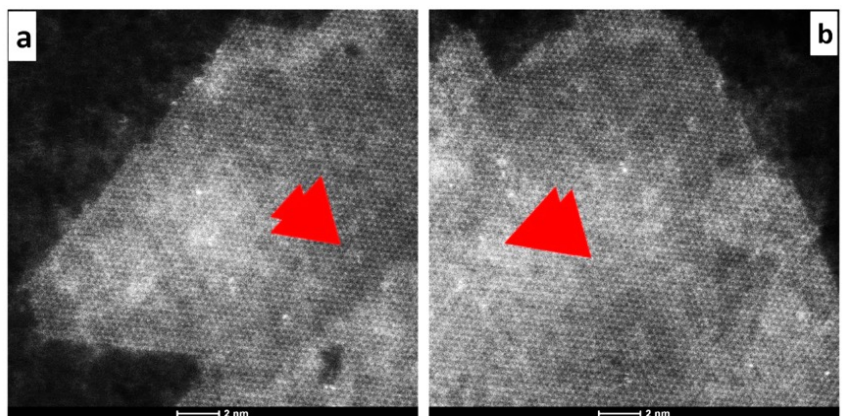

Figure S9. ADF-STEM image of MoS$_2$ flakes merging into a single-crystal monolayer film free of inversion domain boundaries.

Electron energy loss spectroscopy (EELS) performed on the hBN surface reveals two characteristic peaks at the boron and nitrogen edges (Fig. S10a) which correlate with σ* $sp^2$ bonds and π* bonds [29], establishing the hexagonal honeycomb structure of hBN. Energy-dispersive X-ray spectroscopy mapping (EDS, lower panel of Fig. S10) also confirms the chemical fingerprint of the as-grown heterostructure. EDS elemental maps for nitrogen, sulfur and molybdenum are shown, indicating a uniform distribution of N in the hBN substrate, while Mo and S are locally observed within the triangular domain. All the STEM EDS maps were collected using the superX EDS quad-detectors on Titan[3]. The EDS boron elemental map is inconclusive because of low signal levels due to weak X-ray generation from this low atomic number element.

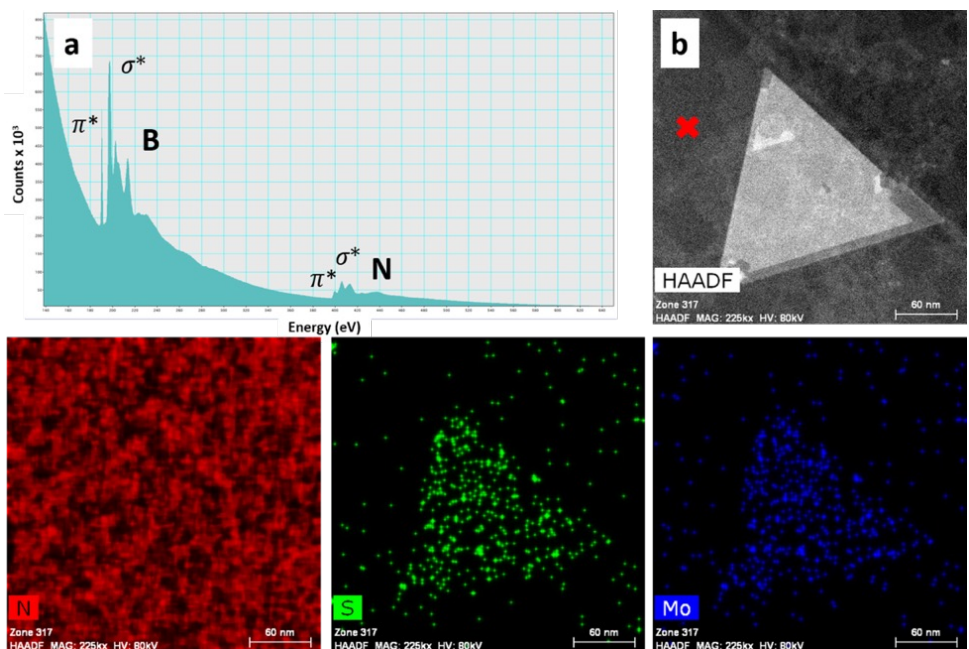

Figure S10. (a) Point EEL spectrum of hBN surface clearly shows the boron and nitrogen edges (point is marked by red cross in panel b). (lower panels) STEM-EDS elemental maps for nitrogen, sulfur, and molybdenum.

**Reactive Ion Etching**

The hBN flakes were placed on Si/SiO$_2$ substrate for reactive ion etching. 50 sccm of oxygen gas was introduced into a Tepla M4L Plasma Etch system operating at a pressure of 200 mTorr, with a radio frequency (13.56 MHz) power of 50W to generate plasma. Different etching times were applied to suspended hBN surfaces while fixing other plasma generation parameters. After plasma treatment, we use the same PMMA transfer method to obtain freestanding hBN substrates for the growth of MoS$_2$/hBN heterostructures.

**Quantifying the degree of epitaxy upon etching**

With increasing etching time, the degree of epitaxy tends to decrease and the size of as-synthesized MoS$_2$ flakes decreases (Fig. S11), since more surface defects and dangling bonds are generated that can act as active nucleation sites for growth of MoS$_2$ flakes. This observation suggests that the relation of defect concentration and degree of orientation control is an interplay of both defect density and defect type: while the number of nucleation sites is increased upon etching, more complex defect structures (e.g. step edges at the edge of triangular voids) do not facilitate epitaxy. Moreover, whereas computational studies suggest that a single defect under a grown region can determine the stacking orientation at early stages of growth (i.e. while the flakes are still nanometer-sized), a higher density of the induced defects could impair this mechanism by interacting strongly with flake edges.

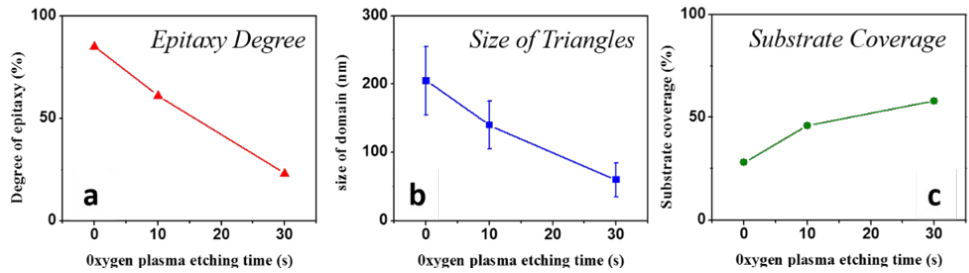

Figure S11. Statistical analysis of (a) the degree of epitaxy, (b) domain size and (c) substrate coverage of the as-grown MoS$_2$ on hBN.

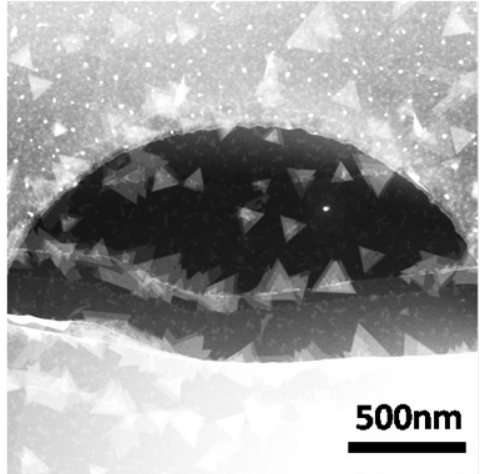

Figure S12. ADF-STEM image of MoS$_2$ domains at step edges (of the pristine hBN without any plasma treatment) shows random orientations due to more complex interactions with the h-BN step.

**Material Characterization**

Scanning electron microscopy was carried out on a Leo 1530 FESEM. Raman/PL characterization was conducted on a 532nm Witec confocal Raman system at an operation power of ~1 mW. AC-S/TEM imaging and spectroscopy were carried out on a FEI Titan$^3$ 60-300 microscope operating at 80kV with a monochromated gun and spherical aberration corrected lenses, providing sub-angstrom resolution. A high angle annular dark field (HAADF) detector was used for the ADF-STEM imaging. The HAADF detector (Fischione) had a collection angle of 51–300 mrad, a beam current of 45pA, and beam convergence angle of 30 mrad (C2 aperture of 70um) for STEM image acquisition. Imaging of surface defects was carried out at the electron dose of ~5000 e$^-$/Å$^2$·s to minimize structural damage [30]. The HREM imaging condition for hBN surface defects was tuned to a negative Cs to provide white atom contrast at a slight over focus.

**Residual strain in the hBN substrate**

Strain can amplify hBN's sublattice asymmetry by e.g. charge redistribution between B and N [31] and uniaxial strain in particular can further lower lattice symmetry. To test the possible role of strain in orientational epitaxy, in Fig. S13 we extended the calculations discussed in Fig. 1 in the main text and calculated the variation of stacking energies for uniformly strained and uniaxially strained (along zigzag or armchair) hBN substrates at a strain of 2% using √21×√21+√13×√13 supercells and DFT-D3 for dispersion forces. This is larger than the typical 0–1% residual strain in bulk hBN reported in literature [32,33] and below the nonlinear elastic threshold [34] of 8%. Fig. S13 shows that the 0/180º near-degeneracy and the rotational barrier persist under strain.

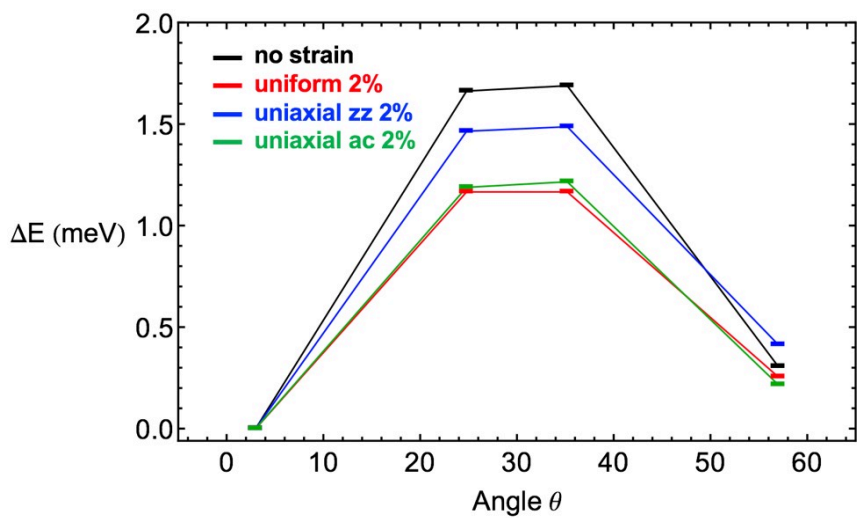

Figure S13. Relative energies (per $MoS_2$ unit) of different stacking orientations for uniformly strained and uniaxially strained hBN substrates at a magnitude of 2%, calculated with DFT-D3 and √21×√21+√13×√13 supercells.

**Strong binding between $V_B$ and precursors other than an isolated Mo atom**

To demonstrate $V_B$ can bind to more realistic metal precursor clusters, we examine the example of $MoO_3$ among the commonly studied $MoO_x$ clusters [35] since the suboxides have oxidation states in between the Mo case (examined in the main text) and the trioxide case (examined here). Spin-polarized DFT calculations show that a $MoO_3$ cluster binds to hBN by 1.5 eV and to $V_B$ by 4.5 eV (Fig. S14), demonstrating the strong likelihood of $V_B$ trapping metal precursors due to its dangling bond character. The detailed structures of $MoO_3$ are not discussed here for reasons mentioned in the previous section on finite flakes.

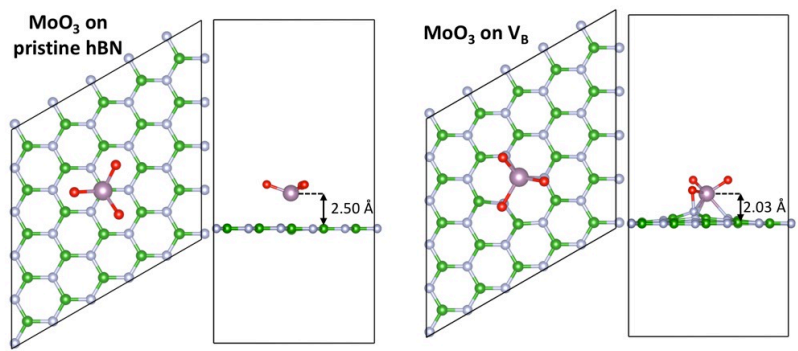

Figure S14. Relaxed structures of MoO$_3$ on pristine hBN and on a boron vacancy.

**Reference**

[1]  J. P. Perdew, K. Burke, and M. Ernzerhof, Phys. Rev. Lett. **77**, 3865 (1996).
[2]  J. P. Perdew, K. Burke, and M. Ernzerhof, Phys. Rev. Lett. **78**, 1396 (1997).
[3]  D. Joubert, Phys. Rev. B **59**, 1758 (1999).
[4]  P. E. Blöchl, Phys. Rev. B **50**, 17953 (1994).
[5]  G. Kresse and J. Furthmüller, Phys. Rev. B **54**, 11169 (1996).
[6]  S. Grimme, J. Antony, S. Ehrlich, and H. Krieg, J. Chem. Phys. **132**, 154104 (2010).
[7]  A. Tkatchenko and M. Scheffler, Phys. Rev. Lett. **102**, 073005 (2009).
[8]  K. Lee, É. D. Murray, L. Kong, B. I. Lundqvist, and D. C. Langreth, Phys. Rev. B **82**, 081101 (2010).
[9]  J. He, K. Hummer, and C. Franchini, Phys. Rev. B **89**, 075409 (2014).
[10] T. Björkman, A. Gulans, A. V. Krasheninnikov, and R. M. Nieminen, J. Phys. Condens. Matter **24**, 424218 (2012).
[11] T. Björkman, A. Gulans, A. V. Krasheninnikov, and R. M. Nieminen, Phys. Rev. Lett. **108**, 235502 (2012).
[12] T. Bučko, S. Lebègue, J. Hafner, and J. G. Ángyán, Phys. Rev. B **87**, 064110 (2013).
[13] J. Heyd, G. E. Scuseria, and M. Ernzerhof, J. Chem. Phys. **118**, 8207 (2003).
[14] J. Heyd, G. E. Scuseria, and M. Ernzerhof, J. Chem. Phys. **124**, 219906 (2006).
[15] J. Kibsgaard, T. F. Jaramillo, and F. Besenbacher, Nat. Chem. **6**, 248 (2014).
[16] B. Huang, H. Xiang, J. Yu, and S. H. Wei, Phys. Rev. Lett. **108**, 206802 (2012).
[17] H.-P. Komsa and A. V. Krasheninnikov, Phys. Rev. B **91**, 125304 (2015).
[18] C. Freysoldt, B. Grabowski, T. Hickel, J. Neugebauer, G. Kresse, A. Janotti, and C. G. Van De Walle, Rev. Mod. Phys. **86**, 253 (2014).
[19] C. Attaccalite, M. Bockstedte, A. Marini, A. Rubio, and L. Wirtz, Phys. Rev. B **83**, 144115 (2011).

[20] W. Orellana and H. Chacham, Phys. Rev. B **63**, 125205 (2001).
[21] F. Zhang, C. Erb, L. Runkle, X. Zhang, and N. Alem, Nanotechnology **29**, 025602 (2018).
[22] F. Zhang, M. A. AlSaud, M. Hainey, K. Wang, J. M. Redwing, and N. Alem, Microsc. Microanal. **22**, 1640 (2016).
[23] F. Zhang, K. Momeni, M. A. AlSaud, A. Azizi, M. F. Hainey, J. M. Redwing, L. Q. Chen, and N. Alem, 2D Mater. **4**, 025029 (2017).
[24] H. Yu, Z. Yang, L. Du, J. Zhang, J. Shi, W. Chen, P. Chen, M. Liao, J. Zhao, J. Meng, G. Wang, J. Zhu, R. Yang, D. Shi, L. Gu, and G. Zhang, Small **13**, 1603005 (2017).
[25] M. Bosi, RSC Adv. **5**, 75500 (2015).
[26] C. Lee, H. Yan, L. E. Brus, T. F. Heinz, J. Hone, and S. Ryu, ACS Nano **4**, 2695 (2010).
[27] A. Splendiani, L. Sun, Y. Zhang, T. Li, J. Kim, C. Y. Chim, G. Galli, and F. Wang, Nano Lett **10**, 1271 (2010).
[28] K. F. Mak, C. Lee, J. Hone, J. Shan, and T. F. Heinz, Phys. Rev. Lett. **105**, 136805 (2010).
[29] N. Alem, Q. M. Ramasse, C. R. Seabourne, O. V. Yazyev, K. Erickson, M. C. Sarahan, C. Kisielowski, A. J. Scott, S. G. Louie, and A. Zettl, Phys. Rev. Lett. **109**, 205502 (2012).
[30] J. C. Meyer, A. Chuvilin, G. Algara-Siller, J. Biskupek, and U. Kaiser, Nano Lett. **9**, 2683 (2009).
[31] M. Neek-Amal, J. Beheshtian, A. Sadeghi, K. H. Michel, and F. M. Peeters, J. Phys. Chem. C **117**, 13261 (2013).
[32] W. Yang, G. Chen, Z. Shi, C.-C. Liu, L. Zhang, G. Xie, M. Cheng, D. Wang, R. Yang, D. Shi, K. Watanabe, T. Taniguchi, Y. Yao, Y. Zhang, and G. Zhang, Nat. Mater. **12**, 792 (2013).
[33] A. Summerfield, A. Davies, T. S. Cheng, V. V. Korolkov, Y. Cho, C. J. Mellor, C. T. Foxon, A. N. Khlobystov, K. Watanabe, T. Taniguchi, L. Eaves, S. V. Novikov, and P. H. Beton, Sci. Rep. **6**, 22440 (2016).
[34] J. Wu, B. Wang, Y. Wei, R. Yang, and M. Dresselhaus, Mater. Res. Lett. **1**, 200 (2013).
[35] J. A. Oliveira, W. B. De Almeida, and H. A. Duarte, Chem. Phys. Lett. **372**, 650 (2003).